\documentstyle[preprint,tighten,aps,epsfig]{revtex}
\begin{document} 
\preprint{MKPH-T-99-10}
\draft
\title{Generalized polarizabilities of the pion in chiral perturbation theory}
\author{C.\ Unkmeir,$^1$ S.\ Scherer,$^1$ 
A.\ I.\ L'vov,$^2$ D.\ Drechsel$^1$}
\address{$^1$ Institut f\"ur Kernphysik, Johannes Gutenberg-Universit\"at,
J.\ J.\ Becher-Weg 45, D-55099 Mainz, Germany}
\address{$^2$ P.N. Lebedev Physical Institute, Moscow, 117924, Russia} 
\date{April 23, 1999}
\maketitle
\begin{abstract}
   We present a calculation of the virtual Compton scattering amplitude for
$\gamma^\ast+\pi\to \gamma+\pi$ in the framework of chiral perturbation 
theory at ${\cal O}(p^4)$.
   We explicitly derive expressions for generalized electromagnetic 
polarizabilities and discuss alternative def\/initions of these
quantities.
\end{abstract}

\section{Introduction}
   Compton scattering of real photons (RCS) is one of the simplest reactions
for obtaining information on the structure of a stable composite system.
   When expanded in the frequency of the photon, the leading-order term
of the low-energy scattering amplitude is specif\/ied by the model-independent
Thomson limit in terms of the charge and the mass of the target.
   Genuine structure effects f\/irst appear at second order 
and can be parametrized in terms of 
the electric and magnetic polarizabilities
(for an overview see, e.g., Refs.\ 
\cite{Holstein_90,Lvov_93,Scherer_99}).
   As there is no stable pion target, the empirical information on the 
electromagnetic polarizabilities has been extracted from 
high-energy pion-nucleus bremsstrahlung 
\cite{Antipov_83,Antipov_85} and radiative pion photoproduction
off the nucleon \cite{Aibergenov_86}.
   In principle, the electromagnetic polarizabilities of the 
pion also enter into the crossed process $\gamma\gamma\to\pi\pi$. 
   However, there is some debate concerning the accuracy of extracting
these quantities from the crossed channel 
\cite{Babusci_92,Kaloshin_92,Donoghue_93,Kaloshin_94,Portoles_95}.

   From a theoretical point of view, a precise determination of the
pion polarizabilities is of great importance, since (approximate) chiral 
symmetry allows one to predict the electromagnetic polarizabilities 
of the charged pion in terms
of the radiative decay $\pi^+\to e^+\nu_e\gamma$ \cite{Terentev_73}.
   Corrections to the leading-order PCAC result have been calculated
at ${\cal O}(p^6)$ in chiral perturbation theory and turn out to
be be rather small \cite{Buergi_96}.
   New experiments are presently being carried out \cite{Ahrens_95}
or have been proposed \cite{Moinester_97,Gorringe_98} 
to signif\/icantly reduce the uncertainties in the empirical results 
and thus subject the predictions of chiral symmetry to a stringent 
test.

   Clearly, the possibilities to investigate the structure of the target 
increase substantially if virtual photons are used, because energy and 
three-momentum can be varied independently and, furthermore,
the longitudinal component of the transition current can be explored. 
   In particular, virtual Compton scattering (VCS) off the nucleon, 
as tested in the reaction $e^-+p\to e^- + p +\gamma$, has attracted 
considerable interest (see, e.g., Refs.\ \cite{dHose_97,Guichon_98}).
   The pion-VCS amplitude of $\gamma^\ast+\pi\to \gamma+\pi$ can, 
in principle, be studied through the inelastic scattering of high-energy
pions off atomic electrons,
$\pi+e^-\to\pi+e^-+\gamma$.
   Such events are presently analyzed as part of the SELEX E781 experiment
\cite{Moinester_99}.

   In this paper, we will investigate the VCS reaction  
$\gamma^\ast+\pi\to \gamma+\pi$ in the framework of 
chiral perturbation theory at ${\cal O}(p^4)$.
   We will f\/irst give a short survey of chiral perturbation theory
and then def\/ine our conventions for the VCS invariant amplitude.
   We then discuss the result for the soft-photon and residual amplitudes, 
respectively.
   Finally, the model-dependent residual amplitude is analyzed in terms
of alternative def\/initions of generalized polarizabilities.

\section{The chiral Lagrangian}
   Chiral perturbation theory (ChPT) \cite{Weinberg_79,Gasser_84,Gasser_85} 
is based on the chiral $\mbox{SU(2)}_L\times \mbox{SU(2)}_R$ symmetry
of QCD in the limit of vanishing $u$- and $d$-quark masses. 
   The assumption of spontaneous symmetry breaking down to $\mbox{SU(2)}_V$ 
gives rise to three massless pseudoscalar Goldstone bosons with vanishing 
interactions in the limit of zero energies.
   These Goldstone bosons are identif\/ied with the physical pion triplet,
the nonzero pion masses resulting from an explicit symmetry breaking
in QCD through the quark masses.
   The effective Lagrangian of the pion interaction is organized in a 
so-called momentum expansion,
\begin{equation}  
  {\cal L}_{\mbox{\footnotesize eff}}
={\cal L}_2+{\cal L}_4+\cdots, 
\end{equation}
   where the subscripts refer to the order in the expansion.
   Interactions with external f\/ields, such as the electromagnetic
f\/ield, as well as explicit symmetry breaking due to the f\/inite
quark masses, are systematically incorporated into the effective
Lagrangian.
   Covariant derivatives and quark-mass terms count as ${\cal O}(p)$ 
and ${\cal O}(p^2)$, respectively.
   Weinberg's power counting scheme \cite{Weinberg_79} allows for a 
classif\/ication of the Feynman diagrams by establishing a relation between 
the momentum expansion and the loop expansion.
    The most general chiral Lagrangian at ${\cal O}(p^2)$ is given by
\begin{equation}
\label{l2}
{\cal L}_2 = \frac{F^2}{4} \mbox{Tr} \left[ D_{\mu} U (D^{\mu}U)^{\dagger} 
+\chi U^{\dagger}+ U \chi^{\dagger} \right],
\end{equation}
where $U$ is a unimodular unitary $(2\times 2)$ matrix, transforming as
$V_R U V_L^\dagger$ for $(V_L,V_R)\in \mbox{SU(2)}_L\times\mbox{SU(2)}_R$.
   As a parametrization of $U$ we will use   
\begin{equation}
\label{paru}
U(x)=\frac{\sigma(x)+i\vec{\tau}\cdot\vec{\pi}(x)}{F}, 
\quad \sigma^2(x)+\vec{\pi}^2(x)=F^2,
\end{equation}
where $F$ denotes the pion-decay constant in the chiral limit: 
$F_\pi=F[1+{\cal O}(\hat{m})]=92.4$ MeV. 
   We will work in the isospin-symmetric limit $m_u=m_d=\hat{m}$. 
   The quark mass is contained in $\chi=2 B_0 \hat{m}=m^2_\pi$
at ${\cal O}(p^2)$, where $B_0$ is related to the quark condensate
$<\!\!\bar{q}q\!\!>$.
   The covariant derivative $D_\mu U = \partial_\mu U +\frac{i}{2}e A_\mu 
[\tau_3,U]$ contains the coupling to the electromagnetic f\/ield $A_\mu$.
   The most general structure of ${\cal L}_4$, f\/irst obtained by 
Gasser and Leutwyler (see Eq.\ (5.5) of Ref.\ \cite{Gasser_84}), reads,
in the standard trace notation,
\begin{eqnarray}
\label{l4gl}
{\cal L}^{GL}_4 &=&
\frac{l_1}{4} \left\{\mbox{Tr}[D_{\mu}U (D^{\mu}U)^{\dagger}] \right\}^2
+\frac{l_2}{4}\mbox{Tr}[D_{\mu}U (D_{\nu}U)^{\dagger}]
\mbox{Tr}[D^{\mu}U (D^{\nu}U)^{\dagger}]
+\frac{l_3}{16}\left[\mbox{Tr}(\chi U^\dagger+ U\chi^\dagger)\right]^2
\nonumber\\
&&+\frac{l_4}{4}\mbox{Tr}[D_\mu U(D^\mu\chi)^\dagger
+D_\mu\chi(D^\mu U)^\dagger]
+l_5\left[\mbox{Tr}(F^R_{\mu\nu}U F^{\mu\nu}_LU^\dagger)
-\frac{1}{2}\mbox{Tr}(F_{\mu\nu}^L F^{\mu\nu}_L
+F_{\mu\nu}^R F^{\mu\nu}_R)\right]\nonumber\\
&&+i\frac{l_6}{2}\mbox{Tr}[ F^R_{\mu\nu} D^{\mu} U (D^{\nu} U)^{\dagger}
+ F^L_{\mu\nu} (D^{\mu} U)^{\dagger} D^{\nu} U]
-\frac{l_7}{16}\left[\mbox{Tr}(\chi U^\dagger-U\chi^\dagger)\right]^2
+\cdots,
\end{eqnarray}
   where three terms containing only external f\/ields have been omitted.
   For the electromagnetic interaction, the f\/ield-strength tensors  
are given by $F^{\mu\nu}_L=F^{\mu\nu}_R=-\frac{e}{2}\tau_3 
(\partial^\mu A^\nu-\partial^\nu A^\mu)$.

\section{Conventions}
\begin{figure}
\begin{center}
\epsfig{file=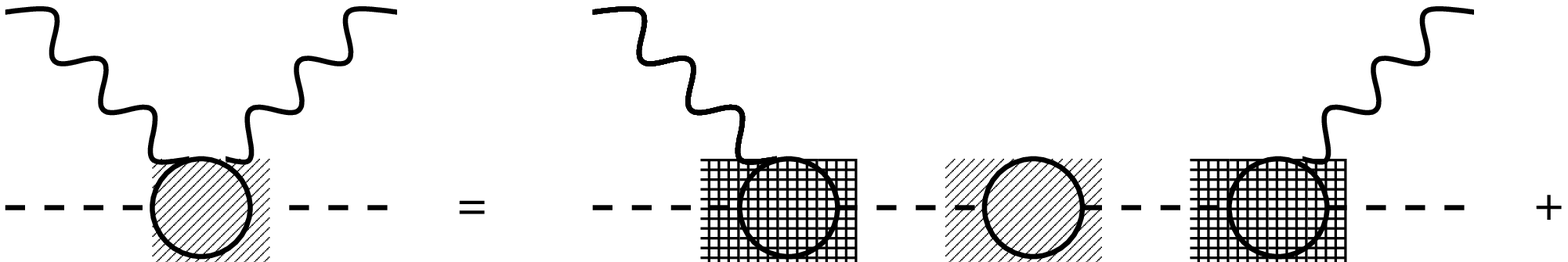,width=14cm}
\caption{\label{vcsdiagrams.fig}
Compton scattering amplitude:
Hatched and cross-hatched vertices denote one-particle-reducible
and one-particle-irreducible contributions, respectively.
All building blocks are renormalized.
}
\end{center}
\end{figure}

   In the following, we will discuss the VCS amplitude for 
$\gamma^\ast(q,\epsilon)+\pi^i(p_i)\to\gamma(q',\epsilon')+\pi^j(p_f)$ 
($q^2\leq 0$, $q'^2=0$, $q'\cdot\epsilon'=0$).
   Throughout the calculation we use the conventions of Bjorken and Drell 
\cite{Bjorken_1964} with $e^2/4\pi\approx1/137$, $e>0$. 
   For the isospin decomposition of the invariant amplitude we use
\begin{equation}
\label{isospin}
{\cal M}_{ij}=\delta_{ij}{\cal A}+(\delta_{ij}-\delta_{i3}\delta_{j3})
{\cal B},
\end{equation}
   where $i$ and $j$ denote the {\em cartesian} isospin indices of the
initial and f\/inal pions, respectively.
   With the def\/inition 
\begin{displaymath}
|\pi^\pm(p)\!>=\frac{1}{\sqrt{2}}[a_1^\dagger(p)\pm i a_2^\dagger(p)]|0\!>,
\quad
|\pi^0(p)\!>=a_3^\dagger(p)|0\!>,
\end{displaymath}
we may express the physical amplitudes in terms of the isospin amplitudes
\begin{eqnarray}
\label{mcharged}
{\cal M}_{\pi^+}={\cal M}_{\pi^-}&=&\frac{1}{2}({\cal M}_{11}+{\cal M}_{22})
={\cal A}+{\cal B},\\
\label{mneutral}
{\cal M}_{\pi^0}&=&{\cal M}_{33}={\cal A}.
\end{eqnarray}
   We split the contributions to ${\cal M}_{ij}$ into a pole piece ($P$) and 
a one-particle-irreducible, residual part ($R$), 
${\cal A}={\cal A}_P+{\cal A}_R$,
${\cal B}={\cal B}_P+{\cal B}_R$
(see Fig.\ \ref{vcsdiagrams.fig}).
   Since the $\pi^0$ is its own antiparticle, the electromagnetic vertex
$\pi^0\pi^0\gamma^\ast$ vanishes due to charge-conjugation invariance
and hence ${\cal A}_P\equiv 0$.
   In general, the pole piece ${\cal B}_P$ and the one-particle-irreducible 
piece ${\cal B}_R$ are not separately gauge invariant.

\section{Soft-photon amplitude}
   According to Weinberg's power counting, a calculation of the $s$- and 
$u$-channel pole terms at ${\cal O}(p^4)$ involves the renormalized 
irreducible vertex at ${\cal O}(p^4)$, 
\begin{equation}
\label{gammamu}
\Gamma^\mu(p',p)=(p'+p)^\mu F(q^2)+(p'-p)^\mu \frac{p'^2-p^2}{q^2}[
1-F(q^2)],\quad q=p'-p,
\end{equation}
   where $F(q^2)$ is the prediction for the electromagnetic form factor
of the pion (see Eq.\ (15.3) of Ref.\ \cite{Gasser_84}).
   To that order, the renormalized propagator is simply given by
\begin{equation}
\label{prop}
i\Delta_R(p)=\frac{i}{p^2-m_\pi^2+i0^+},
\end{equation}
with $m^2_\pi$ the ${\cal O}(p^4)$ result for the pion mass squared 
(see Eq.\ (12.2) of Ref.\ \cite{Gasser_84}). 
   Note that Eqs.\ (\ref{gammamu}) and (\ref{prop}) satisfy the 
Ward-Takahashi identity \cite{Ward_50,Takahashi_57} 
$q_\mu \Gamma^\mu(p',p)=\Delta_R^{-1}(p')-\Delta_R^{-1}(p)$.
 
\begin{figure}
\begin{center}
\epsfig{file=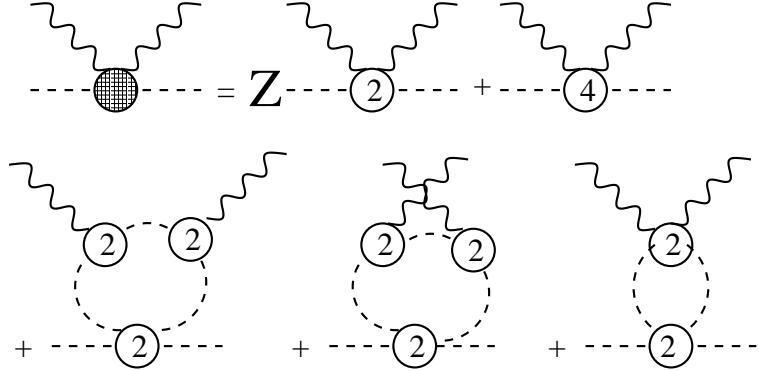,width=10cm}
\caption{\label{classb.fig}
   Diagrams contributing to the one-particle-irreducible residual amplitude 
at ${\cal O}(p^4)$. 
   Vertices derived from ${\cal L}_{2n}$ are denoted by $2n$ in the
interaction blobs.
   $Z$ denotes the wave function renormalization factor corresponding
to the Lagrangian of Eq.\ (\ref{l2}) and the pion f\/ield of Eq.\ 
(\ref{paru}).
   At ${\cal O}(p^4)$, only the contribution from ${\cal L}_2$ has to be
multiplied by $Z$.}
\end{center}
\end{figure}

   With these ingredients the result for ${\cal B}_P$ at ${\cal O}(p^4)$ reads
\begin{equation}
\label{bp}
{\cal B}_P=-ie^2\left\{F(q^2)\left[\frac{2p_f\cdot\epsilon'^\ast\,
(2p_i+q)\cdot\epsilon}{s-m_\pi^2}
+\frac{(2p_f-q)\cdot\epsilon\, 2p_i\cdot\epsilon'^\ast}{u-m_\pi^2}\right]
+2 q\cdot\epsilon\, q\cdot\epsilon'^\ast \frac{1-F(q^2)}{q^2}\right\},
\end{equation}
   which is easily seen not to be gauge invariant by itself. 
   The set of one-particle-irreducible diagrams is shown in 
Fig.\ \ref{classb.fig} and gives rise to a residual part of the form
\begin{equation}
\label{br}
{\cal B}_R=ie^2\left[2\epsilon\cdot\epsilon'^\ast
+2(q^2\epsilon\cdot\epsilon'^\ast-q\cdot \epsilon\,
q\cdot\epsilon'^\ast)\frac{F(q^2)-1}{q^2}\right]+\tilde{\cal B}_R.
\end{equation}
   We combine Eqs.\ (\ref{bp}) and (\ref{br}) into the form
\begin{equation}
\label{b}
{\cal B}=\tilde{\cal B}_P+\tilde{\cal B}_R,
\end{equation}
where
\begin{equation}
\label{tildebp}
\tilde{\cal B}_P=-ie^2 F(q^2)\left[\frac{2p_f\cdot\epsilon'^\ast\,
(2p_i+q)\cdot\epsilon}{s-m_\pi^2}
+\frac{(2p_f-q)\cdot\epsilon\, 2p_i\cdot\epsilon'^\ast}{u-m_\pi^2}
-2\epsilon\cdot\epsilon'^\ast\right],
\end{equation}
   with the result that $\tilde{\cal B}_P$ and $\tilde{\cal B}_R$ 
are now separately gauge invariant.
   In particular, $\tilde{\cal B}_P$ has the form of the soft-photon
result obtained in Eq.\ (10) of Ref.\ \cite{Fearing_98}.
   A somewhat different approach for obtaining the soft-photon
result can be found in Ref.\ \cite{Lvov_99}.

\section{Residual amplitudes}
    As has been discussed in detail in Ref.\ \cite{Drechsel_1997},
a gauge-invariant parametrization of the residual amplitude for 
$\gamma^\ast+\pi\to \gamma+\pi$ can be written in terms of three 
invariant functions $f_i(q^2,q\cdot q',q\cdot P)$, where $P=p_i+p_f$.
   At ${\cal O}(p^4)$, the result for the residual isospin amplitudes 
${\cal A}_R$ and $\tilde{\cal B}_R$ reads:
\begin{eqnarray}
\label{ar}
{\cal A}_R&=&-ie^2(q'\cdot\epsilon\, q\cdot \epsilon'^\ast-q\cdot q'
\epsilon\cdot\epsilon'^\ast)
\frac{m^2_\pi+2q\cdot q'-q^2}{8\pi^2F_\pi^2 q\cdot q'}
{\cal G}(q^2,q\cdot q'),\\
\label{tbr}
\tilde{\cal B}_R&=&-ie^2(q'\cdot\epsilon \,q\cdot \epsilon'^\ast-q\cdot q'
\epsilon\cdot\epsilon'^\ast)
\left[-\frac{4(2 l_5^r-l_6^r)}{F^2_\pi}
-\frac{2 m^2_\pi+2q\cdot q'-q^2}{16\pi^2 F^2_\pi q\cdot q'}
{\cal G}(q^2,q\cdot q')\right],
\end{eqnarray}
   where the combination $2l^r_5-l^r_6=(2.85\pm 0.42)\times 10^{-3}$ is
determined  through the decay $\pi^+\to e^+\nu_e\gamma$.
   In Eqs.\ (\ref{ar}) and (\ref{tbr}) we have introduced the abbreviation
\begin{equation}
\label{functiong}
{\cal G}(q^2,q\cdot q')=
1+\frac{m^2_\pi}{q\cdot q'}\left[J^{(-1)}(a)-J^{(-1)}(b)\right]
-\frac{q^2}{2q\cdot q'}\left[J^{(0)}(a)-J^{(0)}(b)\right],
\end{equation}
where 
\begin{displaymath}
J^{(n)}(x):=\int_0^1 dy y^n\ln[1+x(y^2-y)-i0^+]
\end{displaymath}
and 
\begin{displaymath}
a:=\frac{q^2}{m^2_\pi},\quad b:=\frac{q^2-2q\cdot q'}{m^2_\pi}.
\end{displaymath}
   The one-loop integrals $J^{(0)}$ and $J^{(-1)}$ are given by
(see Appendix C of Ref.\ \cite{Bellucci_1994}\footnote{In reproducing
these results we found Refs.\ \cite{Barbieri_1972} and
\cite{tHooft_1979} useful.})
\begin{eqnarray*}
J^{(0)}(x)
&=& \left \{ \begin{array}{l}
-2-\sigma\ln\left(\frac{\sigma-1}{\sigma+1}\right)\quad(x<0),\\
-2+2\sqrt{\frac{4}{x}-1}\,\mbox{arccot}
\left(\sqrt{\frac{4}{x}-1}\right)\quad (0\le x<4),\\
-2-\sigma\ln\left(\frac{1-\sigma}{1+\sigma}\right)-i\pi\sigma
\quad(4<x),
\end{array} \right.\\
J^{(-1)}(x)&=&\left \{ \begin{array}{l}
\frac{1}{2}\ln^2\left(\frac{\sigma-1}{\sigma+1}\right)\quad(x< 0),\\
-\frac{1}{2}\arccos^2\left(1-\frac{x}{2}\right)\quad (0\le x <4),\\
\frac{1}{2}\ln^2\left(\frac{1-\sigma}{1+\sigma}\right)-\frac{\pi^2}{2}
+i\pi\ln\left(\frac{1-\sigma}{1+\sigma}\right)\quad (4< x),
\end{array} \right.
\end{eqnarray*}
with 
\begin{displaymath}
\sigma(x)=\sqrt{1-\frac{4}{x}},\quad x\notin [0,4].
\end{displaymath}
   A comparison of Eqs.\ (\ref{ar}) and (\ref{tbr}) with Eq.\ (18) of Ref.\ 
\cite{Drechsel_1997} shows that, at 
${\cal O}(p^4)$, only one of the three functions 
$f_i(q^2,q\cdot q',q\cdot P)$ contributes, 
i.e., $f_2=f_3=0$. 
   Furthermore, at this order in the chiral expansion, the function $f_1$ 
does not depend on $q\cdot P=q'\cdot P$. 
   Our result for ${\cal A}_R$ is in agreement with Ref.\ 
\cite{Belkov_97}, where the photoproduction of neutral pion pairs in
the Coulomb f\/ield of a nucleus was studied.

\section{Generalized polarizabilities of Guichon, Liu, and Thomas}
   In order to discuss the generalized polarizabilities, we expand 
the function ${\cal G}$ of Eq.\ (\ref{functiong}) for negative $q^2$ 
around $q\cdot q'=0$,
\begin{equation}
\label{functiongexp}
{\cal G}(q^2,q\cdot q')=-\frac{q\cdot q'}{m^2_\pi} 
J^{(0)'}\left(\frac{q^2}{m^2_\pi}\right)+
{\cal O}[(q\cdot q')^2],\quad J^{(0)'}(x)=\frac{d J^{(0)}(x)}{dx},
\end{equation}
where
\begin{equation}
\label{jop}
J^{(0)'}(x)=\frac{1}{x}\left[1-\frac{2}{x\sigma}
\ln\left(\frac{\sigma-1}{\sigma+1}\right)\right]
=\frac{1}{x}\left[1+2J^{(-1)'}(x)\right],\quad x< 0.
\end{equation}
   For the charged and neutral pion we obtain, respectively,
\begin{eqnarray}
\label{f1pp}
f_1^{\pi^\pm}(q^2,q\cdot q',q\cdot P)&=&-\frac{4(2l^r_5-l_6^r)}{F^2_\pi}
+\frac{2q\cdot q'-q^2}{16\pi^2 F^2_\pi q\cdot q'}
{\cal G}(q^2,q\cdot q')\nonumber\\
&=&-\frac{4(2l^r_5-l_6^r)}{F^2_\pi}+\frac{q^2}{16\pi^2 F^2_\pi m^2_\pi}
J^{(0)'}\left(\frac{q^2}{m^2_\pi}\right)
+{\cal O}(q\cdot q'),\\
\label{f1p0}
f_1^{\pi^0}(q^2,q\cdot q',q\cdot P)&=&
-\frac{1}{8\pi^2 F^2_\pi}\left(1-\frac{q^2}{m^2_\pi}\right)
J^{(0)'}\left(\frac{q^2}{m^2_\pi}\right)
+{\cal O}(q\cdot q').
\end{eqnarray}

   We will f\/irst discuss the generalized polarizabilities as def\/ined
in Ref.\ \cite{Guichon_1995}, where the residual amplitude was analyzed in 
the photon-pion center-of-mass frame in terms of a multipole expansion.
   Only terms {\em linear} in the frequency of the f\/inal photon were kept,
and the result was parametrized in terms of ``generalized polarizabilities.''
   The connection with the covariant approach was established in Ref.\
\cite{Drechsel_1997}, where it was also found that only two of the three 
polarizabilities $P^{(01,01)0}$, $P^{(11,11)0}$, and $\hat{P}^{(01,1)0}$
of Ref.\ \cite{Guichon_1995} are independent, once 
the constraints due to charge conjugation are combined with 
particle-crossing symmetry. 
   According to Eqs.\ (35) and (36) of Ref.\ \cite{Drechsel_1997}
we def\/ine generalized electric and magnetic polarizabilities 
$\alpha(|\vec{q}|^2)$ and $\beta(|\vec{q}|^2)$, respectively, as
\begin{eqnarray}
\label{alpha1}
\alpha(|\vec{q}|^2)&\equiv&
-\frac{e^2}{4\pi}\sqrt{\frac{3}{2}}P^{(01,01)0}(|\vec{q}|)\nonumber\\
&=&
\frac{e^2}{8\pi m_\pi}\sqrt{\frac{m_\pi}{E_i}}
\left[-f_1(\omega_0^2-|\vec{q}|^2,0,0)+2m_\pi\frac{|\vec{q}|^2}{\omega_0}
f_2(\omega_0^2-|\vec{q}|^2,0,0)\right],\\
\label{beta1}
\beta(|\vec{q}|^2)&\equiv&-\frac{e^2}{4\pi}\sqrt{\frac{3}{8}}P^{(11,11)0}
(|\vec{q}|)=
\frac{e^2}{8\pi m_\pi}\sqrt{\frac{m_\pi}{E_i}}
f_1(\omega_0^2-|\vec{q}|^2,0,0),
\end{eqnarray}
   where $\omega_0=q_0|_{\omega'=0}=m_\pi-\sqrt{m_\pi^2+|\vec{q}|^2}$.
   A few remarks are in order at this point.
\begin{enumerate}
\item In our present work, we strictly stick to the convention 
of Ref.\ \cite{Bjorken_1964}. 
   This is why Eqs.\ (\ref{alpha1}) and (\ref{beta1}) differ by an overall
factor $1/2m_\pi$ from Ref.\ \cite{Drechsel_1997}, where in
Eq.\ (1) an additional factor $2m_\pi$ was introduced for the
spin-0 case.
\item The variable $q^2$ only appears in the combination
$q^2/m^2_\pi$, resulting in 
\begin{displaymath}
\left.\frac{q^2}{m^2_\pi}\right|_{\omega'=0}=2\frac{m_\pi-E_i}{m_\pi},\quad
E_i=\sqrt{m_\pi^2+|\vec{q}|^2}.
\end{displaymath}
\item The factor $\sqrt{m_\pi/E_i}$ originates from
an additional normalization factor ${\cal N}$ in Eq.\ (32) of Ref.\
\cite{Guichon_1995}, such that
\begin{displaymath}
\frac{2m_\pi}{\sqrt{4 E_i E_f}}\stackrel{\omega'\to 0}{\to}
\sqrt{\frac{m_\pi}{E_i}}.
\end{displaymath}
\end{enumerate}
   Using the results of Eqs.\ (\ref{f1pp}) and (\ref{f1p0}) together
with $f_2=0$, we then obtain
\begin{eqnarray}
\label{alphapp1}
\alpha_{\pi^\pm}(|\vec{q}|^2)&=&-\beta_{\pi^\pm}(|\vec{q}|^2)
=
\frac{e^2}{8\pi m_\pi}\sqrt{\frac{m_\pi}{E_i}}
\left[\frac{4(2l^r_5-l^r_6)}{F^2_\pi}-2 \frac{m_\pi-E_i}{m_\pi}
\frac{1}{(4\pi F_\pi)^2} J^{(0)'}\left(2\frac{m_\pi-E_i}{m_\pi}\right)\right],
\nonumber\\
&&\\
\label{alphap01}
\alpha_{\pi^0}(|\vec{q}|^2)&=&-\beta_{\pi^0}(|\vec{q}|^2)
=\frac{e^2}{4\pi} \frac{1}{(4\pi F_\pi)^2 m_\pi}\sqrt{\frac{m_\pi}{E_i}}
\left(1-2\frac{m_\pi- E_i}{m_\pi}\right)
J^{(0)'}\left(2\frac{m_\pi- E_i}{m_\pi}\right).
\end{eqnarray}
   At the one-loop level, the $|\vec{q}|^2$ dependence is entirely
given in terms of the pion mass $m_\pi$ and the pion-decay
constant $F_\pi$, i.e., no additional ${\cal O}(p^4)$ low-energy
constant enters.
   At $|\vec{q}|^2=0$, Eqs.\ (\ref{alphapp1}) and (\ref{alphap01}) 
reduce to the RCS polarizabilities \cite{Holstein_90}
\begin{eqnarray}
\bar{\alpha}_{\pi^\pm}&=&-\bar{\beta}_{\pi^\pm}=\frac{e^2}{4\pi} 
\frac{2}{m_\pi F^2_\pi}(2l^r_5-l^r_6)
=(2.68\pm0.42)\times 10^{-4}\,\mbox{fm}^3,\\
\bar{\alpha}_{\pi^0}&=&-\bar{\beta}_{\pi^0}
=-\frac{e^2}{4\pi}\frac{1}{96\pi^2 F^2_\pi m_\pi}
=-0.50 \times 10^{-4}\,\mbox{fm}^3,
\end{eqnarray}
where we made use of $J^{(0)'}(0)=-\frac{1}{6}$.
   At ${\cal O}(p^6)$, the RCS predictions for the
charged pion read
$\bar{\alpha}_{\pi\pm}=(2.4\pm 0.5)\times 10^{-4}\,\mbox{fm}^3$ and 
$\bar{\beta}_{\pi^\pm}=(-2.1\pm 0.5)\times 10^{-4}\,\mbox{fm}^3$
\cite{Buergi_96}.
   The corresponding corrections amount to a 12\% (24\%) change
of the ${\cal O}(p^4)$ result, indicating a good convergence.
   We also note that the original degeneracy $\bar{\alpha}=-\bar{\beta}$
is lifted at ${\cal O}(p^6)$.
   The predictions of ChPT have to be compared with the empirical
results $\bar{\alpha}_{\pi\pm}=(6.8\pm 1.4)\times 10^{-4}\,\mbox{fm}^3$ 
\cite{Antipov_83}, 
$\bar{\alpha}_{\pi\pm}=(20\pm 12)\times 10^{-4}\,\mbox{fm}^3$ 
\cite{Aibergenov_86},
and 
$\bar{\beta}_{\pi^\pm}=(-7.1\pm 4.6)\times 10^{-4}\,\mbox{fm}^3$
\cite{Antipov_85}.
   Clearly, an improved accuracy is required to test the chiral
predictions.
   For the neutral pion, the ${\cal O}(p^6)$ corrections turn out to
be much larger, 
$\bar{\alpha}_{\pi^0}=(-0.35\pm 0.10)\times 10^{-4}\,\mbox{fm}^3$ and
$\bar{\beta}_{\pi^0}=(1.50\pm 0.20)\times 10^{-4}\,\mbox{fm}^3$
\cite{Bellucci_1994}.

\section{Alternative def\/inition of the generalized dipole polarizabilities}

   Another generalization of the RCS polarizabilities
is obtained by parametrizing the invariant amplitude as\footnote{ 
A detailed discussion will be given in Ref.\ \cite{Lvov_99}.}
\begin{equation}
\label{mvcs2}
-i{\cal M}=B_1 F^{\mu\nu} F'_{\mu\nu} 
+\frac{1}{4}B_2(P_\mu F^{\mu\nu})(P^\rho F'_{\rho\nu})
+\frac{1}{4}B_5(P^\nu q^\mu F_{\mu\nu})(P^\sigma q^\rho F'_{\rho\sigma}),
\end{equation}
   where $F^{\mu\nu}$ and $F'_{\mu\nu}$ refer, respectively, to 
the gauge-invariant combinations
\begin{displaymath}
F^{\mu\nu}=-i q^\mu \epsilon^\nu+iq^\nu \epsilon^\mu,\quad
F'_{\mu\nu}=iq'_\mu\epsilon'^\ast_\nu-iq'_\nu\epsilon'^\ast_\mu.
\end{displaymath}
   The functions $B_1$, $B_2$, and $B_5$ are {\em even} functions of $P$.
   Introducing the suggestive notation 
$$\vec{E}=i(q_0\vec{\epsilon}-\vec{q}\epsilon_0),\quad
\vec{B}=i\vec{q}\times\vec{\epsilon},\quad
\vec{E}'=-i(q_0'\vec{\epsilon}\,'^\ast-\vec{q}\,'\epsilon_0'^\ast),\quad
\vec{B}'=-i\vec{q}\,'\times\vec{\epsilon}\,'^\ast,
$$
  the structures of Eq.\ (\ref{mvcs2}) are particularly simple when
evaluated in the pion Breit frame (p.B.f.) def\/ined by $\vec{P}=0$, 
\begin{eqnarray*}
F^{\mu\nu} F'_{\mu\nu}&=&[-2\vec{E}\cdot\vec{E}'+2\vec{B}\cdot\vec{B}']_{
p.B.f.},\\
P_\mu F^{\mu\nu} P^\rho F'_{\rho\nu}&=&
[-P^2_0 \vec{E}\cdot\vec{E}']_{p.B.f.},\\
P^\nu q^\mu F_{\mu\nu} P^\rho q^\sigma F'_{\sigma\rho}&=&
[P^2_0 \vec{q}\cdot \vec{E}\vec{q}\cdot\vec{E}']_{p.B.f.}.
\end{eqnarray*}
   Note that by def\/inition $[P^2_0]_{p.B.f.}=P^2$.
   In the p.B.f., Eq.\ (\ref{mvcs2}) can thus be expressed as
\begin{equation}
\label{mvcsnbf}
-i{\cal M}=\left[2 B_1 \vec{B}\cdot\vec{B}' 
-\left(2B_1+\frac{P^2}{4} B_2\right)
\vec{E}\cdot\vec{E}'
+\frac{P^2}{4}B_5 \vec{q}\cdot \vec{E} \vec{q}\cdot\vec{E}'\right]_{p.B.f.}.
\end{equation}   
      Since $\vec{E}=\vec{E}_T+\vec{E}_L$, $\vec{E}\cdot\vec{E}'$ contains
both transverse and longitudinal components with respect to $\hat{q}$,
for which reason we will introduce the quantities $\alpha_T$ and $\alpha_L$ below:
\begin{equation}
\label{mvcsnbf2}
-i{\cal M}=\left\{
2 B_1 \vec{B}\cdot\vec{B}' -\left(2B_1+\frac{P^2}{4} B_2\right)
\vec{E}_T\cdot\vec{E}'
+\left[\frac{P^2}{4}B_5 |\vec{q}|^2-\left(2B_1+\frac{P^2}{4} B_2\right)\right]
\vec{E}_L\cdot\vec{E}'\right\}_{p.B.f.}.
\end{equation} 
   We now consider the limit $\omega'\to 0$ of the residual amplitudes,
for which $B_i^r\to b_i^r(q^2)$, and def\/ine three generalized dipole 
polarizabilities in terms of the invariants of Eq.\ (\ref{mvcs2}),
\begin{eqnarray}
\label{beta}
8\pi m_\pi\beta(q^2)&\equiv&2b_1^r(q^2),\\
\label{alphat}
8\pi m_\pi
\alpha_T(q^2)&\equiv&-2b_1^r(q^2)-\left(M^2-\frac{q^2}{4}\right)b_2^r(q^2),\\
\label{alphal}
8\pi m_\pi\alpha_L(q^2)&\equiv&-2b_1^r(q^2)-\left(M^2-\frac{q^2}{4}
\right)[b_2^r(q^2)+q^2 b_5^r(q^2)],
\end{eqnarray}
   the superscript $r$ referring to the residual amplitudes beyond the 
soft-photon result.  
   In general, the transverse and longitudinal electric polarizabilities
$\alpha_T$ and $\alpha_L$ will differ by a term, vanishing however in the 
RCS limit $q^2=0$.
   Comparing with Eq.\ (\ref{mvcsnbf2}), 
the generalized dipole polarizabilities are seen to be def\/ined such
that they multiply the structures $\vec{B}\cdot\vec{B}'$,
$\vec{E}_T\cdot\vec{E}'$, and $\vec{E}_L\cdot\vec{E}'$, respectively,
as $\omega'\to 0$.
   We note that $[\vec{B}\cdot\vec{B}']_{p.B.f.}$ and $
[\vec{E}_L\cdot \vec{E}']_{p.B.f.}$ are of ${\cal O}(\omega')$
whereas $[\vec{E}_T\cdot\vec{E}']_{p.B.f.}= {\cal O}(\omega'^2)$, 
i.e., that different powers of $\omega'$ have been kept.

   At $q^2=0$, the usual RCS polarizabilities are recovered,
\begin{equation}
\label{rcslimit}
\beta(0)=\bar{\beta},\quad
\alpha_L(0)=\alpha_T(0)=\bar{\alpha}.
\end{equation}
   The connection to the generalized polarizabilities of Guichon {\em et al.}
\cite{Guichon_1995} can either be established by direct comparison 
or via the results of Ref.\ \cite{Drechsel_1997},
\begin{eqnarray}
\alpha(|\vec{q}|^2)&=&-\frac{e^2}{4\pi} \sqrt{\frac{3}{2}}P^{(01,01)0}(
|\vec{q}|)
=\sqrt{\frac{m_\pi}{E_i}}\alpha_L(\omega_0^2-\vec{q}\,^2),\\
\beta(|\vec{q}|^2)&=&-\frac{e^2}{4\pi} \sqrt{\frac{3}{8}}P^{(11,11)0}(
|\vec{q}|)
=\sqrt{\frac{m_\pi}{E_i}}\beta(\omega_0^2-\vec{q}\,^2),
\end{eqnarray}
   with $\omega_0=q_0|_{\omega'=0}=m_\pi-E_i$, 
$\omega_0^2-\vec{q}\,^2=2m_\pi(m_\pi-E_i)$, and $E_i=
\sqrt{m^2_\pi+\vec{q}\,^2}$, and all variables referring to the cm frame.
   Using Eqs.\ (35) - (37) of Ref.\ \cite{Drechsel_1997}, we f\/ind 
that the transverse electric dipole polarizability is part of a 
second-order contribution in $\omega'$ beyond the approximation
of Guichon {\em et al.}, 
\begin{equation}
\alpha_T(q^2)=\alpha_L(q^2)+\frac{e^2}{4\pi}(4M^2-q^2)q^2
\tilde{f}_3(q^2,0,0),
\end{equation}
   where $q\cdot P \tilde{f}_3\equiv f_3$.

\begin{figure}
\begin{center}
\epsfig{file=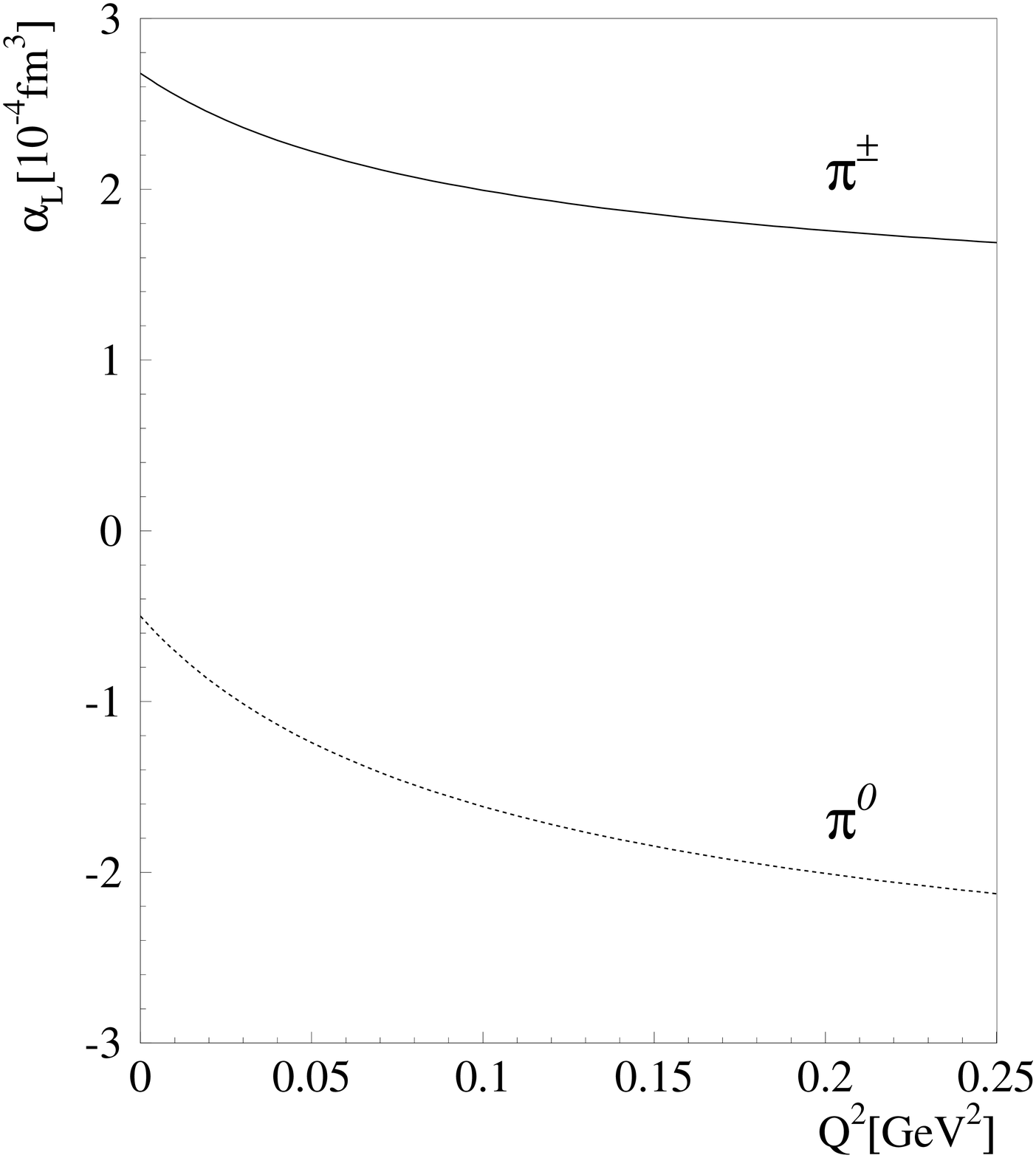,width=10cm}
\caption{\label{alpha.fig}
${\cal O}(p^4)$ prediction for the generalized dipole polarizabilities
$\alpha_L(-Q^2)$ of the charged pion (solid curve) and the neutral pion
(dashed curve) as function of $Q^2$ 
[see Eqs.\ (\ref{alphapp2}) and (\ref{alphap02})]. 
   At ${\cal O}(p^4)$,  $\alpha_L(q^2)=\alpha_T(q^2)=-
\beta(q^2)$.
}
\end{center}
\end{figure}

   At ${\cal O}(p^4)$, $f_2=f_3=0$, with the result of particularly 
simple expressions for the generalized dipole polarizabilities,
\begin{eqnarray}
\label{alphapp2}
\alpha^{\pi^\pm}_L(q^2)&=&
\alpha^{\pi^\pm}_T(q^2)=
-\beta^{\pi^\pm}(q^2)
=
\frac{e^2}{8\pi m_\pi}
\left[\frac{4(2l^r_5-l^r_6)}{F^2_\pi}-\frac{q^2}{m_\pi^2}
\frac{1}{(4\pi F_\pi)^2} J^{(0)'}\left(\frac{q^2}{m_\pi^2}\right)\right],
\\
\label{alphap02}
\alpha^{\pi^0}_L(q^2)&=&
\alpha^{\pi^0}_T(q^2)=
-\beta^{\pi^0}(q^2)
=\frac{e^2}{4\pi}\frac{1}{(4\pi F_\pi)^2m_\pi}
\left(1-\frac{q^2}{m^2_\pi}\right)
J^{(0)'}\left(\frac{q^2}{m^2_\pi}\right).
\end{eqnarray}
   The results for the generalized dipole polarizabilities are shown
in F\/ig.\ \ref{alpha.fig}.
   Even though chiral perturbation theory is only applicable for 
small external momenta, for the sake of completeness we also quote 
the asymptotic behavior as $q^2\to-\infty$,
\begin{eqnarray}
\label{asbpp}
\alpha^{\pi^\pm}_L(q^2) &\to& \bar{\alpha}_{\pi^\pm}+3\bar{\alpha}_{\pi^0}
=1.18\times 10^{-4}\,\mbox{fm}^3,\\
\alpha^{\pi^0}_L(q^2)&\to& 6 \bar{\alpha}_{\pi^0}
=-3.0\times 10^{-4}\,\mbox{fm}^3.
\end{eqnarray}  
   As in the case of real Compton scattering, we expect the degeneracy
$\alpha_L(q^2)=\alpha_T(q^2)=-\beta(q^2)$ to be lifted at the
two-loop level.

\section{Summary}
   We have calculated the invariant amplitudes for virtual Compton scattering
off the pion, $\gamma^\ast+\pi\to \gamma +\pi$, at the one-loop level, 
${\cal O}(p^4)$, in chiral perturbation theory.
   For the charged pion, the result may be decomposed into a gauge-invariant 
soft-photon amplitude involving the electromagnetic form factor of the pion
and a gauge-invariant residual amplitude.
   For the neutral pion, the soft-photon amplitude vanishes.   
   We have analyzed the low-energy behavior of the residual amplitudes in 
terms of generalized polarizabilities.
   In this context we have introduced two alternative def\/initions of the
generalized polarizabilities, a f\/irst one based on a multipole expansion 
in the center-of-mass frame, and a second one based on a covariant approach 
interpreted in the pion Breit frame.
   The connection between the different approaches has been established.
   In the framework of ChPT at ${\cal O}(p^4)$, the momentum dependence
of the generalized polarizabilities is entirely predicted in terms of the
pion mass and the pion-decay constant, i.e., no additional 
counter-term contribution appears.
   As in the case of real Compton scattering, the results at 
${\cal O}(p^4)$ show a degeneracy of the polarizabilities, 
$\alpha_L(q^2)=\alpha_T(q^2)=-\beta(q^2)$, which we expect to be lifted at 
the two-loop level.

\section{Acknowledgements}
   This work was supported by the Deutsche Forschungsgemeinschaft (SFB 443).
   A.\ L.\ thanks the theory group of the Institut f\"{u}r Kernphysik 
for the hospitality and support during his stay in Mainz where part of 
his work was done.

\end{document}